\documentclass[a4paper,fleqn]{cas-dc}
\usepackage[numbers,sort&compress]{natbib}
\usepackage{multirow,slashed,hyperref,cleveref,oplotsymbl,pifont}
\hypersetup{colorlinks=true, linkcolor=Maroon, citecolor=WildStrawberry, filecolor=WildStrawberry, urlcolor=ForestGreen}

\definecolor{mpl-green}{HTML}{008000}

\newcommand{\ie}{\textit{i.e.}}

\newcommand{\madgraph}{{\sc MG5\_aMC}\xspace}
\newcommand{\delphes}{{\sc Del\-phes\,3}\xspace}
\newcommand{\madanalysis}{{\sc Mad\-A\-na\-ly\-sis\,5}\xspace}
\newcommand{\pythia}{{\sc Pythia\,8}\xspace}
\newcommand{\spey}{{\sc Spey}\xspace}
\newcommand{\pyhf}{{\sc pyhf}\xspace}

\newcommand{\Plus}{\mathord{\text{\ding{58}}}}

\begin{document}

\def\floatpagepagefraction{1}
\def\textpagefraction{.001}

\title[mode = title]{Monojets reveal overlapping excesses for light compressed higgsinos}        \let\printorcid\relax             

\author{Diyar Agin}
\author{Benjamin Fuks}
\author{Mark D. Goodsell}
\author{Taylor Murphy}

\address{
 Laboratoire de Physique Th\'{e}orique et Hautes \'{E}nergies (LPTHE), UMR 7589,\\ Sorbonne Universit\'{e} \& CNRS, 4 place Jussieu, 75252 Paris Cedex 05, France
}

\shortauthors{D.~Agin {\it et~al.}}
\shorttitle{{Monojets reveal overlapping excesses for light compressed higgsinos}}

\begin{abstract}
The ATLAS and CMS collaborations have recently presented results of searches for compressed electroweakinos in final states including soft leptons. These searches are sensitive to mass splittings ranging from quite small values of about 5~GeV to $\mathcal{O}(10)$ GeV, which are endemic to scenarios with wino-like and higgsino-like lightest supersymmetric particles (LSPs). While all experimental results exhibit apparently compatible mild excesses, these soft-lepton analyses, taken together with disappearing-track searches targeting much smaller splittings, notably leave unconstrained a sizeable region of parameter space with modest splittings of \mbox{1--5~GeV}. We point out that this gap can be closed, for scenarios with a higgsino-like LSP, by monojet searches. On the other hand, we find at the same time that current monojet searches show excesses in a region partially overlapping that favoured by the soft-lepton analyses. We provide an up-to-date map of these results and show, among others, a best-fit point with an excess greater than $2\sigma$ that is consistent with a higgsino-like LSP mass around 177~GeV. We finally comment on how such a point can be realised in the MSSM.
\end{abstract}


\maketitle

\section{Introduction}
\label{introduction}

It is often stated that there are no signs of low-energy supersymmetry at the LHC. In particular, new colourful particles are strongly constrained with limits typically in the TeV range (see for example \cite{ATLAS:2020syg, ATLAS-CONF-2020-002, Sirunyan:2019xwh, Sirunyan:2019ctn}). Searches for electroweak production of new particles naturally have much lower reach, but there are now examples where masses above 1~TeV can be excluded \cite{ATLAS:2021yqv}. Such strong bounds apply for winos, which are electroweak triplets thus featuring a large production cross section, in scenarios with a largely split mass spectrum hence yielding large missing momentum or transverse momenta for jets or leptons. In addition, bounds of $\mathcal{O}(1)$~TeV have been imposed in the case of extremely small mass splittings by searches for a charged particle with a long lifetime that leaves a track in the detector~\cite{ATLAS:2019gqq,ATLAS:2022pib} (which may also disappear~\cite{CMS:2020atg,Goodsell:2021iwc,Araz:2021akd,CMS:2023mny}). 

The fermionic sectors of typical supersymmetric models contain higgsinos and at least one bino in addition to the winos; the former are electroweak doublets and the latter singlets. The production of these is suppressed relative to winos, and it has therefore been a challenge to place limits on them. Nevertheless, searches targeting winos can constrain higgsino and bino masses up to $900$ GeV when the spectrum exhibits large mass splitting ($>\!450$ GeV in the case of~\cite{ATLAS:2021yqv}), or even up to $950$~GeV when considering a disappearing track signature~\cite{CMS:2023mny}. On the other hand, \emph{realistic} supersymmetric scenarios typically have non-trivial mixing among the binos, higgsinos and winos, even when one or more of the multiplets are rather heavy. In these cases a spectrum with modest splittings amongst the charged and neutral components, of \mbox{1--10~GeV}, is quite typical~\cite{Giudice:1998xp, Randall:1998uk, Baer:2011ec, Nagata:2014wma, Baer:2015tva}. This is exactly the situation that is most hard to probe.

There has therefore been interest in new searches targeting compressed spectra, and at the same time an effort to extract the most information possible from the data by combining multiple analyses. In particular, the ATLAS collaboration has presented results for compressed electroweakinos from a combination \cite{ATLAS:2021moa} of a three-lepton plus missing-energy ($3\ell + E_{\text{T}}^{\text{miss}}$) channel and a previously studied $2\ell + E_{\text{T}}^{\text{miss}}$ channel \cite{ATLAS:2019lng}, while the CMS collaboration has combined results \cite{CMS:2023qhl} from the $2\ell/3\ell + E_{\text{T}}^{\text{miss}}$ \cite{CMS:2021edw} and $\geq\! 3\ell$ \cite{CMS:2021cox} channels (among others). In doing so, both collaborations found excesses in their respective soft dilepton channels. In an effort to paint a comprehensive picture, they interpret the results in representative wino-like and higgsino-like scenarios with few parameters; for a higgsino-like LSP, the excesses appear for chargino-LSP splittings of roughly \mbox{5--15~GeV}. Moreover, when the soft-lepton search results are viewed together with the most recent disappearing-track analyses, there exists in addition to the soft-lepton excesses a gap in the parameter space, for modest mass splittings of a few GeV, where neither higgsinos nor winos are excluded at any mass above the longstanding LEP limit of roughly 93~GeV \cite{LEPlimits}. It is this gap that motivated the present work.

In this letter we show that the most recent monojet searches \cite{ATLAS:2021kxv,CMS:2021far}, of the type typically used to target stable neutral particles, can close the small-mass-splitting gap. We focus on the higgsino case, as that is the one of most current interest given that it has been hardest to pin down. Surprisingly, however, even as we use these monojet searches to close the gap left open by the soft-lepton channels currently in excess, we find that they also have excesses \emph{in the same parameter space}.

\section{Analysis details and signal simulation}

We focus on the higgsino scenario considered by the ATLAS and CMS collaborations in~\cite{ATLAS:2021moa, CMS:2021edw}, in which the mass splitting between the lightest neutralino ($\tilde{\chi}_1^0$, henceforth the LSP) and next-to-lightest neutralino ($\tilde{\chi}_2^0$) is denoted by $\Delta m$, and where the lightest chargino ($\tilde{\chi}_1^\pm$) has a mass halfway between the two:
\begin{align}\label{eq:split}
    m_{\tilde{\chi}_1^\pm} =  \frac{1}{2}\,(m_{\tilde{\chi}_2^0} + m_{\tilde{\chi}_1^0}).
\end{align}
While ATLAS and CMS simply impose this condition in simplified models with zero mixing between the higgsinos and gauginos, it reflects the nearly symmetrical neutralino-chargino splittings found in realistic models in which the lightest electroweakinos are mostly higgsino and in which small splittings are generated by correspondingly small mixings with the bino and/or wino \cite{PhysRevD.37.2515}. Further discussion of realistic scenarios is offered in our analysis below. The combined CMS analysis~\cite{CMS:2023qhl}, which collects results from several searches in multi-lepton channels in order to cover a range of $\Delta m$ comparable to the ATLAS combined analysis~\cite{ATLAS:2021moa}, instead fixes $m_{\tilde{\chi}^{\pm}_1} = m_{\tilde{\chi}^0_2}$, a benchmark motivated by scenarios with a bino-like $\tilde{\chi}^0_1$ and wino-like $\tilde{\chi}^{\pm}_1$ and $\tilde{\chi}^0_2$. Thus only the component analyses \cite{CMS:2021edw, CMS:2021cox} can be interpreted within the scenario defined by \eqref{eq:split}, and in this letter we are unable to display results from both the ATLAS and CMS combined soft-lepton analyses together. We therefore consider the exclusion region\footnote{The limits have been provided on \href{https://www.hepdata.net/record/ins1866951}{\textsc{HEPData}}.} of the combined ATLAS soft-lepton analysis~\cite{ATLAS:2021moa}, alongside that of the CMS $2\ell/3\ell$ analysis \cite{CMS:2021edw}, and complement those with the results of the latest CMS disappearing-track search~\cite{CMS:2023mny} since these depend only on the chargino-LSP mass splitting.

We combine this set of searches with a recast of the ATLAS monojet search~\cite{ATLAS:2021kxv} and the monojet regions of the CMS monojet/mono-$V$ search~\cite{CMS:2021far}, both of which are available on the Public Analysis Database (PAD) of \madanalysis~\cite{Dumont:2014tja, Conte:2018vmg, DVN/IRF7ZL_2021, DVN/REPAMM_2023}. It has been pointed out that higgsinos are either barely or not at all constrained by monojet searches. If we take a pure higgsino configuration in which all higgsino states are degenerate, then this is true because the charged components are stable and so would be vetoed in the signal event selections. In this case only the process $pp \rightarrow \tilde{\chi}_1^0 \tilde{\chi}_2^0$ therefore contributes. But such a parameter point is excluded by searches exploiting the presence of an isolated charged tracks up to large higgsino masses anyway. Instead, as soon as we consider the (more realistic) case of non-degenerate higgsinos, the charged components can decay to the LSP, and will therefore also contribute to a jet plus missing energy signature.

Naively, the efficiency of a search for an energetic jet and large missing transverse momentum should not depend much on how the produced particles decay, and it might therefore be expected that the constraints would be roughly independent of $\Delta m$. However, the decays of the $\tilde{\chi}_2^0$ and $\tilde{\chi}_1^\pm$ higgsinos proceed through off-shell $Z$ and $W$ bosons, and so lead to (somewhat soft) jets or leptons. For a typical monojet search, this behaviour makes it harder to discriminate signals from background events, which come from \emph{on-shell} $Z$-boson and $W$-boson production. Hence the two analyses we consider implement selections on both jets and leptons: the ATLAS search requires large missing energy ($>200$~GeV), and at least one energetic jet with transverse momentum $p_{\text{T}} > 150$~GeV, but at most four signal jets and no leptons; the CMS analysis requires a minimum transverse energy of $100$~GeV for the leading jet, no isolated sufficiently hard leptons ($p_{\text{T}} > 10$~GeV), as well as requirements on the minimum angle between any jets and the missing momentum (to avoid fake missing momentum from miscalculated angles) which may be violated for signals with more jets. Hence in our model, as $\Delta m$ increases, we will rapidly produce sufficient hard leptons for our signal to be rejected by the cuts, and the search loses sensitivity. The exact degree to which this occurs can, however, only be judged by conducting a study based on a full simulation of the signal.

To perform this study for compressed higgsinos, we first use \textsc{MadGraph5\texttt{\textunderscore}aMC@NLO} (\madgraph) \cite{Alwall:2014hca} to generate hard events describing the production of $\tilde{\chi}_1^0 \tilde{\chi}_2^0$, $\tilde{\chi}_i^0 \tilde{\chi}_1^\pm$, and $\tilde{\chi}_1^+ \tilde{\chi}_1^-$ pairs with up to two additional hard jets. Leading-order (LO) matrix element calculations rely on the UFO~\cite{Darme:2023jdn} model available from~\cite{Duhr:2011se}, and a convolution with the LO set of NNPDF4.0 parton densities~\cite{NNPDF:2021njg} driven through \textsc{LHAPDF}~\cite{Buckley:2014ana}. We next use \pythia~\cite{Sjostrand:2014zea} to perform MLM matching~\cite{Mangano:2006rw, Alwall:2008qv}, parton showering and hadronisation. The detector simulation is performed by the \madanalysis Simplified Fast Detector Simulation (SFS) module~\cite{Araz:2020lnp} for the ATLAS analysis, and for the CMS analysis by \delphes~\cite{deFavereau:2013fsa} with a detector parametrisation provided by the CMS collaboration~\cite{DVN/IRF7ZL_2021}. Signal cross sections are normalised to include next-to-leading order corrections matched with threshold resummation at the next-to-leading-logarithmic accuracy (NLO+NLL), which we compute using the {\sc Resummino} package~\cite{Fuks:2013vua,Fiaschi:2023tkq}. Statistical analysis, including the calculation of expected and observed limits at 95\% confidence level~\cite{Read:2002hq}, is performed after the analysis of the generated event samples with \madanalysis with the aid of the package {\sc Spey}~\cite{Araz:2023bwx}. Finally, in order to investigate the specific realistic supersymmetric scenario discussed below, we also make some preliminary use of a recast of the ATLAS off-shell soft-lepton analyses \cite{ATLAS:2019lng,ATLAS:2021moa}, as well as the two monojet searches, within the package {\sc HackAnalysis} \cite{Goodsell:2021iwc}, details of which will appear elsewhere (we welcome private communication concerning the code in the meantime).

\begin{figure*}
  \centering
  \includegraphics[width=0.97\textwidth]{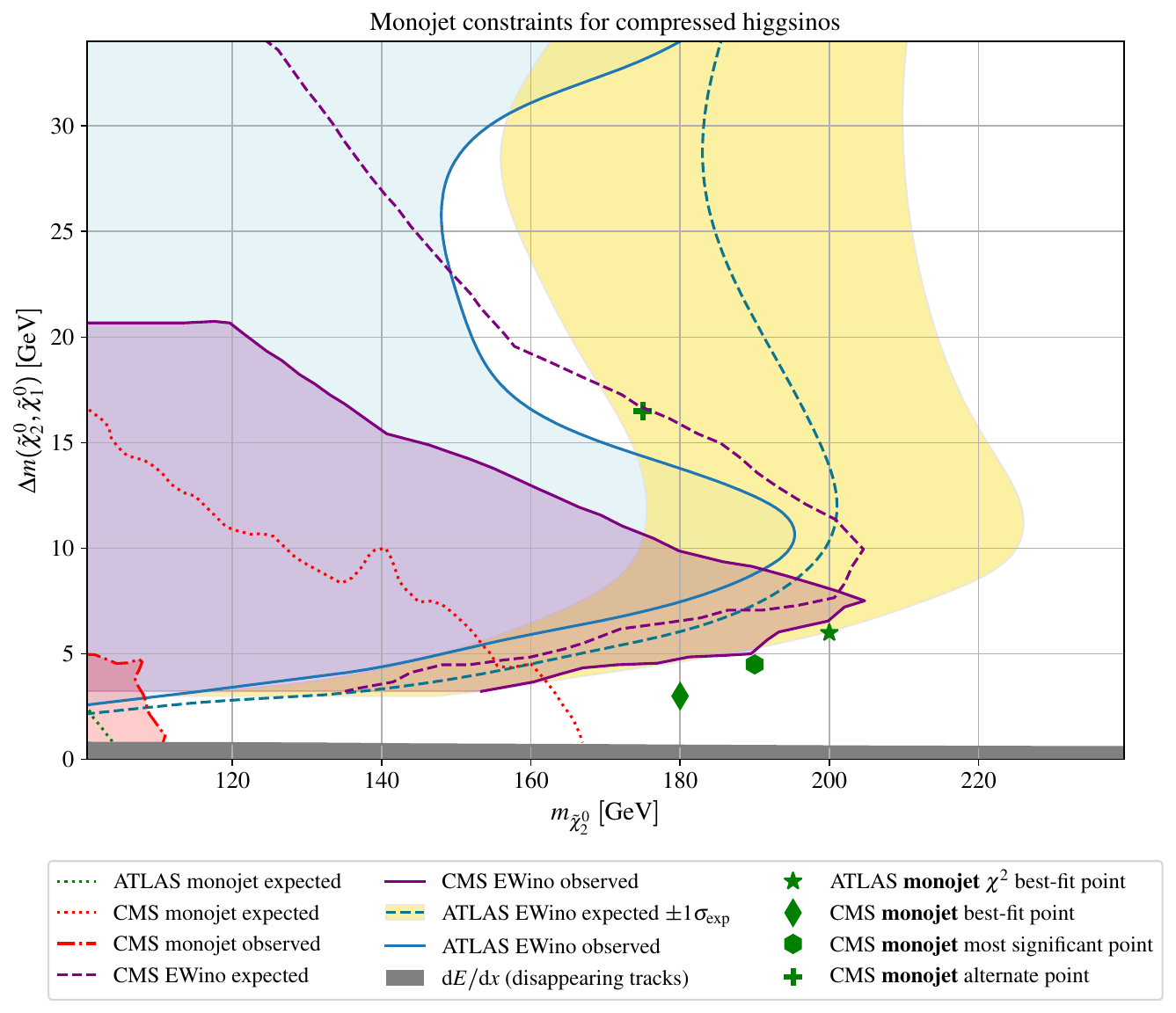}
 \caption{\label{FIG:MoneyPlot} Exclusion regions in the $(m_{\tilde{\chi}_2^0},\Delta m)$ plane as obtained through the reinterpretation of the results of several LHC analyses. The new results relative to the experimental state of the art consist of the ATLAS and CMS monojet exclusion curves, together with the best-fit points potentially explaining the observed excesses. We hence show the best-fit value originating from the CMS monojet analysis, and also the point that yields the greatest significance. For the ATLAS monojet search, we show both the excluded region based on the best expected bin, and the `best fit' to the excess based on a naive $\chi^2$ analysis as described in the text. In addition, we show the existing expected and observed bounds that could be imposed from soft-lepton and disappearing-track searches.} 
\end{figure*}

\section{Results}
Monojet constraints on the higgsino model defined by \eqref{eq:split} are determined from a scan in the $(m_{\tilde{\chi}_2^0},\Delta m)$ plane, which is displayed in Figure \ref{FIG:MoneyPlot} and includes both the ATLAS and CMS monojet expected and observed exclusion contours. We overlay these on the existing limits from the ATLAS and CMS soft-lepton analyses (`EWino') and CMS disappearing-track search (`$\text{d}E/\text{d}x$'). As noted above, the CMS soft-lepton limits for the higgsino scenario considered here are only available for the $2\ell/3\ell$ channel, and not for the latest combination. The CMS monojet analysis provides the correlations between signal regions for the background model in a simplified-likelihood framework, and so permits the calculation of an accurate limit combining all signal regions. In contrast, the ATLAS monojet search does not provide any statistical information, which greatly limits the power of the analysis. It must be considered as a set of separate signal bins that cannot be combined, and from which we can thus only derive bounds from the best region based on expected data. As a result, for practical purposes (\ie, above the LEP limit for charginos), only the CMS monojet search is able to partially close the gap between the soft-lepton and disappearing-track searches. But both monojet searches exhibit striking excesses.

We comment that we extend our monojet limits down to the disappearing-track limits, via an interpolation between the degenerate case and limits computed at $1.5$~GeV. To be strict, in the region $\Delta m < 1.5$~GeV the decays into hadrons will form pions directly, and so are difficult to simulate; our results in that region should not be taken exactly. But it is clear that there will be a limit, because the chargino is still sufficiently short-lived to produce missing energy (and the pions produced will lead to fewer jets).

\subsection{Analysis of the overlapping excesses}

We naturally wish to evaluate the compatibility between these new monojet excesses and the existing soft-lepton excesses. The different information made public as part of each analysis, particularly for monojets, forbids a single unified statistical analysis. Here we describe several well motivated treatments and note the differences in their results.

\subsubsection{ATLAS monojet: $\chi^2$ analysis}

In the absence of official information from ATLAS concerning correlations between bins, we first construct a naive $\chi^2$ function of the signal strength $\mu$ for all of the exclusive (non-overlapping) $E_{\text{T}}^{\text{miss}}$ bins in the single signal region of the analysis \cite{ATLAS:2021kxv}, assuming these bins to be uncorrelated. These bins are labelled EM0, EM1, \dots, EM12 and have widths of 50 or 100 GeV, such that for instance the lowest bin EM0 requires $E_{\text{T}}^{\text{miss}} \in (200,250)$~GeV, since $E_{\text{T}}^{\text{miss}} > 200$~GeV is the baseline missing energy selection. The exclusive bins are listed in Table 1 of the ATLAS analysis, along with the inclusive bins used for model-independent interpretation (which we do not include in our $\chi^2$ function) and the rest of the selection details. For completeness we note that the leading jet in this monojet analysis is required to have transverse momentum $p_{\text{T}}(j_1) > 150$~GeV.

Our $\chi^2$ function for the exclusive bins takes the form
\begin{align}
\chi^2 (\mu) = \sum_{i=1}^{N_{\text{bins}}}\frac{1}{\sigma_i^2}\,[n_{i} -(b_i + \mu s_i)]^2\equiv \sum_{i=1}^{N_{\text{bins}}} \tilde{\chi}_i^2 (\mu),
\label{eq:chitilde}\end{align}
where $n_{i}$ is the number of observed events for a given bin $i$, $b_i$ is the number of background events expected for that bin, $s_i$ is the expected number of signal events, and $\sigma_i$ is the uncertainty on the number of background events. We then minimise this to find the signal strength $\mu$, and look for the points that give the smallest $\chi^2$ with values of $\mu$ close to unity. We find a point at $m_{\tilde{\chi}_2^0} = 200$ GeV and $\Delta m = 6$~GeV that yields $\mu_{\rm min} = 1.005$. This point is marked by {\color{mpl-green}$\bigstar$} and labelled `ATLAS monojet $\chi^2$ best-fit point' on Figure~\ref{FIG:MoneyPlot}. This result must be understood with the aforementioned caveat about unknown correlations between the bins, which we expect to be strong. 

\begin{figure}
  \centering
  \includegraphics[width=0.95\columnwidth]{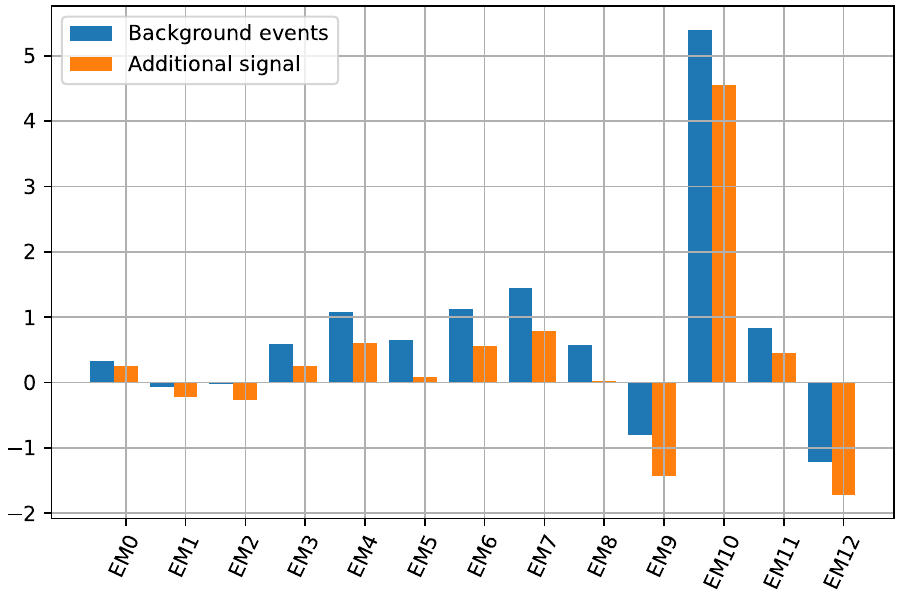}
 \caption{\label{FIG:Hist} Histogram representing the effect of adding our signal events for the calculation of the `best fit' to the excess in the ATLAS monojet search~\cite{ATLAS:2021kxv} based on a naive $\chi^2$ test as described in the text. We display bin-by-bin deviations for exclusive bins as defined by the $\tilde{\chi}_i$ variables of eq.~\eqref{eq:chitilde}; $\tilde{\chi}_i(0)$ is shown in blue, and $\tilde{\chi}_i(\mu_{\rm min})$ in orange.} 
\end{figure}

To further explore this ATLAS best-fit point, in Figure~\ref{FIG:Hist} we evaluate in each exclusive bin $i$ of the ATLAS analysis the functions $\tilde{\chi}_i(0)$ (corresponding to the background-only hypothesis) and $\tilde{\chi_i}(\mu_{\rm min})$ defined in equation (\ref{eq:chitilde}). The strongest local excess exceeds $4\sigma$ and arises in the bin EM10, which requires missing energy $E_{\text{T}}^{\text{miss}} \in (1000,1100)$ GeV. It can be seen that there are several bins with $\sim\!\! 1\sigma$ excesses, where our signal greatly improves the agreement with data.

\subsubsection{CMS monojet: likelihood analysis}

In the CMS analysis, on the other hand, the publication of the signal region correlations permits a more robust assessment of the significance of the excess. In particular, we can obtain the value of the signal strength $\hat{\mu}$ that maximises the \emph{full likelihood} $\mathcal{L}(\mu,\theta)$ assuming Poisson-distributed data and Gaussian nuisances $\theta$, and then compute the $p$-value via \spey using the test statistic $q_0$ \cite{Cowan:2010js}:
\begin{align}\label{eq:q0}
    q_0 \equiv -2 \ln \frac{\mathcal{L}(0,\hat{\hat{\theta}}(0))}{\mathcal{L}(\hat{\mu},\hat{\theta})}\ \ \ \text{for}\ \ \ \hat{\mu} \geq 0.
\end{align}
We find a point at $m_{\tilde{\chi}_2^0} = 180$ GeV, $\Delta m = 3$ GeV that yields $\hat{\mu} = 1.008$ and a $p$-value for the background hypothesis of $p=0.048$; we also find a point at $m_{\tilde{\chi}_2^0} = 190$ GeV, $\Delta m = 4.5$~GeV that yields $\hat{\mu} = 1.368$ and $p=0.019$, \ie\ the CMS analysis by itself shows excesses greater than $2\sigma$. The former point is marked by {\color{mpl-green}$\blacklozenge$} in Figure \ref{FIG:MoneyPlot} (`CMS monojet best-fit point') and the latter point is marked by {\color{mpl-green}\hexagofill} (`CMS monojet most significant point'). To further demonstrate the shape and size of the excess region for this analysis, we include one more point with a $2\sigma$ excess at $m_{\tilde{\chi}_2^0} = 175$~GeV, $\Delta m = 16.5$ GeV (hence $m_{\tilde{\chi}^0_1} = 158.5$~GeV) for which $\hat{\mu} = 1.503$ and $p = 0.042$. This final point is marked by {\color{mpl-green}$\Plus$} (`CMS monojet alternate point').

\subsubsection{Back to ATLAS monojet: exploring correlations}

In light of the differences between the ATLAS and CMS monojet analyses, it is perhaps worthwhile to estimate how the ATLAS monojet best-fit point (computed by $\chi^2$ minimisation) may shift under different assumptions regarding the signal bin correlations. To this end, we construct a likelihood function for the ATLAS monojet exclusive bins assuming (i) no correlations between bins and (ii) strong correlations on the order of those reported for the CMS monojet search. In the latter case, to be precise, since we know the uncertainty $\sigma_i$ associated with each bin, the inverse correlation matrix $\Sigma^{-1}$ should have values $\sigma_i^2$ along the diagonals. If we examine the correlations for the CMS search normalised according to the uncertainty in each bin, we find $(\Sigma^{-1}_{\rm CMS})_{ij}/(\sigma_i \sigma_j) \lesssim 0.4$ for different bins within the same year (and much smaller, but apparently nonzero, correlations for different years). Hence we take as a ``large correlation'' ansatz
\begin{align}
(\Sigma_{\rm Fake\ ATLAS}^{-1})_{ij} \underset{(\text{ii})}{=} \begin{cases}
    \sigma_i^2, & i=j,\\
    0.5\, \sigma_i\sigma_j, & i \neq j.
\end{cases}
\end{align}

With this done, we again use \spey to compute the best-fit $\hat{\mu}$ and $p$-value for each point in our scan. In the uncorrelated case (i), we find that all points have strong $p$-values of $\mathcal{O}(10^{-2})$, due mostly to the large excess in bin EM10 discussed earlier. The best-fit point in this case is at $m_{\tilde{\chi}^0_2} = 165~\text{GeV}$, $\Delta m = 4.5~\text{GeV}$. Here the signal strength is $\hat{\mu} = 1.023$ and the significance is $p= 0.019$. The most significant point in this case, by contrast, is at $m_{\tilde{\chi}^0_2} = 155~\text{GeV}$, $\Delta m = 19.5~\text{GeV}$, with $\hat{\mu} = 1.433$ and $p = 0.016$. In the correlated case (ii), we find severely diminished discovery potential; the best-fit point \emph{not} excluded by either soft-lepton analysis is at $(m_{\tilde{\chi}^0_2},\Delta m) = (180,15)$~GeV, which has a signal strength of $\hat{\mu} = 1.009$ associated with a poor significance of $p = 0.277$. Taken together, the two fictitious scenarios (i) and (ii) allow us to estimate the range of possible best-fit points and significances for the ATLAS monojet analysis; we find that the overlap in the monojet and soft-lepton excesses is fairly robust, though the favoured points migrate as the correlations change; and, importantly, the significance can be better than $98\%$ but diminishes as the correlations become strong.

\subsubsection{How well do the excesses overlap?}

Having shown that models capable of explaining excesses in the soft-lepton searches can also explain excesses in the monojet analyses, and having examined their significances, we next consider the significance of the $2\ell$ excess. The ATLAS collaboration provides on  \href{https://www.hepdata.net/record/ins1866951}{\textsc{HEPData}} full likelihood models for the background fit in the \pyhf format~\cite{Heinrich:2021gyp}, and a set of patches containing the information for several data points, some of which are in the region where an excess is seen. The best of these has $m_{\tilde{\chi}_2^0} = 170$ GeV, $m_{\tilde{\chi}_1^0} = 150$~GeV, which has an expected $\mathrm{CL}_s$ value of $\mathrm{CL}_s^{\text{exp}} = 0.02$ (and would thus be excluded) but an observed value of $\mathrm{CL}_s^{\text{obs}}=0.38$; this point has a $p$-value (computed by \spey using \eqref{eq:q0}) of $p=0.02$, rejecting the null (SM) hypothesis at $98\%$ confidence level. In our data, however, this point does not fit well with the CMS monojet excess, since the best-fit signal strength is at $\hat{\mu} = 1.071$ but the $p$-value is only $p=0.111$. Curiously, this point may fit better with the less-sensitive ATLAS monojet excess: in the fictitious uncorrelated case (i) from above, we find $\hat{\mu} = 1.484$ and $p = 0.028$; when strong correlations are applied for case (ii), both parameters increase modestly to $\hat{\mu} = 1.641$ and $p = 0.087$. Unfortunately, due to the sparseness of the points provided by the ATLAS collaboration, there are no other good candidates to investigate.

\subsection{Comparison with a realistic benchmark}

While the $(m_{\tilde{\chi}^0_2},\Delta m)$ plane mapped in Figure \ref{FIG:MoneyPlot} is inspired by simplified pure-higgsino scenarios, it is possible and illuminating to map the joint soft-lepton/monojet excess region onto somewhat more realistic supersymmetric scenarios in which the LSP is still higgsino-like but does contain some non-negligible gaugino component. These scenarios are worth investigating \emph{prima facie} because pure higgsinos have loop-suppressed mass splittings of $\mathcal{O}(0.1\text{--}1)$ GeV \cite{Thomas:1998wy}: in the case of light higgsinos in the MSSM, the $\mathcal{O}(10)$~GeV splittings considered above \emph{require} higgsino-gaugino mixing. We mention as an example a benchmark scenario in the Minimal Supersymmetric Standard Model in which 
\begin{align}
    \mu = 159~\text{GeV},\ \ M_2 = 270~\text{GeV},\ \ \tan \beta = 10;
\end{align}
and all other relevant parameters, including the bino mass $M_1$, are decoupled. In this scenario, the LSP is predominantly higgsino but does contain an $\mathcal{O}(10)\%$ wino component. We use the \textsc{SPheno} package \cite{Porod:2003um,Porod:2011nf,Staub:2017jnp} to compute the spectrum in this scenario, including one-loop corrections to sfermion masses. The physical light electroweakinos have masses
\begin{align}\hspace*{-.8cm}
  m_{\tilde{\chi}^0_1}     \! = \! 140.7~\text{GeV},\ \
  m_{\tilde{\chi}^{\pm}_1} \! = \! 145.6~\text{GeV},\ \
  m_{\tilde{\chi}^0_2}     \! = \! 170.5~\text{GeV},
\end{align}
and so this benchmark point is located in the vicinity of both the best-fit and the most significant point from the CMS monojet analysis. After simulating and analysing $1.6 \times 10^7$ events at this benchmark point using our {\sc HackAnalysis} recast, we find that this point is allowed by all searches considered here. It also evades constraints from the latest searches in the mono-$V$ channel \cite{ATLAS:2018nda} and the hadronic diboson channel~\cite{ATLAS:2021yqv,Carpenter:2023agq}. From the CMS monojet analysis, furthermore, we find a discovery $p$-value of $p=0.1$ with $\hat{\mu} = 0.92$.

On the other hand, the realistic point actually gives very few events in the soft-lepton searches. We specifically find that it yields only $\mathcal{O}(1)$ events in a few bins for the $2\ell$ search (and always less than one event for the $3\ell$ search); accordingly, the $p$-value for the ATLAS searches is of $\mathcal{O}(0.5)$. This deficit originates from three competing effects. First, all decay channels include not just modes involving $W$ and $Z$ bosons but also Higgs-mediated processes. Next, the decays of the next-to-lightest neutralino $\tilde{\chi}_2^0$ also often proceed via a cascade involving the lightest chargino. Finally, the mass splitting between the lightest chargino and the lightest neutralino is much smaller (only a few GeV) in this point than in the (roughly) equivalent scenario extracted from the ATLAS analysis. This increased compression between $\tilde{\chi}^0_1$ and $\tilde{\chi}^{\pm}_1$ is endemic to models with $\mu \lesssim M_2 \ll M_1$ (\ie, with a decoupled bino but only a moderately heavy wino) \cite{Fuks:2017rio}; in the opposite scenario, with $\mu \lesssim M_1 \ll M_2$, $m_{\tilde{\chi}^{\pm}_1}$ instead drifts closer to $m_{\tilde{\chi}^0_2}$. In any case, our realistic point suggests that spectra differing from the arrangement \eqref{eq:split} assumed by the ATLAS and CMS collaborations require different interpretations and invite further investigation. To carry out such a study, we would have to perform a comprehensive search in the higgsino-gaugino space $(\mu,M_1,M_2)$ in order to find points that can explain the excess; given the computing resources required (since we cannot bias the events at generator level because of the complicated branching ratios), we postpone this task for future work.

\section{Outlook}

The quest to constrain or discover light electroweakinos with small mass splittings will continue through the next runs of the LHC. This task now takes on additional importance given concurrent excesses in the ATLAS and CMS soft-lepton channels and, as we have demonstrated here, in their monojet analyses. We have specifically shown two points with high significance relative to background expectation in the CMS monojet analysis that are compatible with the excess in the latest ATLAS soft-lepton analysis. These points are consistent with an LSP mass lower than 200~GeV and splitting of less than 20~GeV between the lightest neutralinos. 

Given the possible importance of the overlapping excesses, at a minimum it would be very interesting to obtain the full statistical model for the ATLAS monojet search, to be able to compare the excesses more clearly. In future work it should also be possible to clearly compare the soft-lepton scenarios considered in the CMS analysis on the same footing. In this letter, since the combination of CMS analyses was not available, we could not obtain exactly equivalent curves to compare. But with our recast of the ATLAS search, and sufficient computing time, it should be possible to compare excesses in the other scenarios considered by the CMS collaboration. Alternatively, this could be performed rapidly with {\sc SModelS}~\cite{Waltenberger:2016vxp, Alguero:2021dig, MahdiAltakach:2023bdn} once accurate efficiency maps have been obtained, allowing for a global likelihood analysis. Since the initial motivation in this letter was to close the small-$\Delta m$ gap with monojets, we leave this to future work.

While the community awaits new results in these channels, it behooves us to consider more scenarios beyond the Standard Model, perhaps both with and without supersymmetry, that can fit the excesses studied in this letter. These can then be analysed using our recast of the ATLAS soft-lepton searches; however, for wider use in the community, it will also be important to port these to the \madanalysis PAD. This work, and an effort to recast the equivalent CMS searches, is ongoing. 

\section*{Acknowledgments}
\noindent BF, MDG, and TM acknowledge support from Grant ANR-21-CE31-0013, Project DMwithLLPatLHC, from the \emph{Agence Nationale de la Recherche} (ANR), France. We thank Jack Y.~Araz for correspondence about \textsc{Spey}.

\bibliographystyle{JHEP}
\bibliography{literature}

\providecommand{\href}[2]{#2}\begingroup\raggedright\begin{thebibliography}{10}

\bibitem{ATLAS:2020syg}
{\scshape ATLAS} Collaboration, G.~Aad et~al., \emph{{Search for squarks and
  gluinos in final states with jets and missing transverse momentum using 139
  fb$^{-1}$ of $\sqrt{s}$ =13 TeV $pp$ collision data with the ATLAS
  detector}}, \href{http://dx.doi.org/10.1007/JHEP02(2021)143}{\emph{J. High
  Energy Phys.} {\bf 02} (2021) 143},
  [\href{http://arxiv.org/abs/2010.14293}{{\tt 2010.14293}}].

\bibitem{ATLAS-CONF-2020-002}
{\scshape ATLAS} Collaboration, G.~Aad et~al., \emph{{Search for new phenomena
  in final states with large jet multiplicities and missing transverse momentum
  using $ \sqrt{s} $ = 13 TeV proton-proton collisions recorded by ATLAS in Run
  2 of the LHC}}, \href{http://dx.doi.org/10.1007/JHEP10(2020)062}{\emph{J.
  High Energy Phys.} {\bf 10} (2020) 062},
  [\href{http://arxiv.org/abs/2008.06032}{{\tt 2008.06032}}].

\bibitem{Sirunyan:2019xwh}
{\scshape CMS} Collaboration, A.~M. Sirunyan et~al., \emph{{Searches for
  physics beyond the standard model with the $M_\mathrm{T2}$ variable in
  hadronic final states with and without disappearing tracks in proton-proton
  collisions at $\sqrt{s}=$ 13 TeV}},
  \href{http://dx.doi.org/10.1140/epjc/s10052-019-7493-x}{\emph{Eur. Phys. J.
  C} {\bf 80} (2020) 3}, [\href{http://arxiv.org/abs/1909.03460}{{\tt
  1909.03460}}].

\bibitem{Sirunyan:2019ctn}
{\scshape CMS} Collaboration, A.~M. Sirunyan et~al., \emph{{Search for
  supersymmetry in proton-proton collisions at 13 TeV in final states with jets
  and missing transverse momentum}},
  \href{http://dx.doi.org/10.1007/JHEP10(2019)244}{\emph{J. High Energy Phys.}
  {\bf 10} (2019) 244}, [\href{http://arxiv.org/abs/1908.04722}{{\tt
  1908.04722}}].

\bibitem{ATLAS:2021yqv}
{\scshape ATLAS} Collaboration, G.~Aad et~al., \emph{{Search for charginos and
  neutralinos in final states with two boosted hadronically decaying bosons and
  missing transverse momentum in $pp$ collisions at $\sqrt {s}$ = 13\,\,TeV
  with the ATLAS detector}},
  \href{http://dx.doi.org/10.1103/PhysRevD.104.112010}{\emph{Phys. Rev. D} {\bf
  104} (2021) 112010}, [\href{http://arxiv.org/abs/2108.07586}{{\tt
  2108.07586}}].

\bibitem{ATLAS:2019gqq}
{\scshape ATLAS} Collaboration, M.~Aaboud et~al., \emph{{Search for heavy
  charged long-lived particles in the ATLAS detector in 36.1 fb$^{-1}$ of
  proton-proton collision data at $\sqrt{s} = 13$ TeV}},
  \href{http://dx.doi.org/10.1103/PhysRevD.99.092007}{\emph{Phys. Rev. D} {\bf
  99} (2019) 092007}, [\href{http://arxiv.org/abs/1902.01636}{{\tt
  1902.01636}}].

\bibitem{ATLAS:2022pib}
{\scshape ATLAS} Collaboration, G.~Aad et~al., \emph{{Search for heavy,
  long-lived, charged particles with large ionisation energy loss in $pp$
  collisions at $\sqrt{s} = 13~\text{TeV}$ using the ATLAS experiment and the
  full Run 2 dataset}},
  \href{http://dx.doi.org/10.1007/JHEP06(2023)158}{\emph{J. High Energy Phys.}
  {\bf 2306} (2023) 158}, [\href{http://arxiv.org/abs/2205.06013}{{\tt
  2205.06013}}].

\bibitem{CMS:2020atg}
{\scshape CMS} Collaboration, A.~M. Sirunyan et~al., \emph{{Search for
  disappearing tracks in proton-proton collisions at $\sqrt{s} =$ 13 TeV}},
  \href{http://dx.doi.org/10.1016/j.physletb.2020.135502}{\emph{Phys. Lett. B}
  {\bf 806} (2020) 135502}, [\href{http://arxiv.org/abs/2004.05153}{{\tt
  2004.05153}}].

\bibitem{Goodsell:2021iwc}
M.~D. Goodsell and L.~Priya, \emph{{Long dead winos}},
  \href{http://dx.doi.org/10.1140/epjc/s10052-022-10188-1}{\emph{Eur. Phys. J.
  C} {\bf 82} (2022) 235}, [\href{http://arxiv.org/abs/2106.08815}{{\tt
  2106.08815}}].

\bibitem{Araz:2021akd}
J.~Y. Araz, B.~Fuks, M.~D. Goodsell and M.~Utsch, \emph{{Recasting LHC searches
  for long-lived particles with MadAnalysis~5}},
  \href{http://dx.doi.org/10.1140/epjc/s10052-022-10511-w}{\emph{Eur. Phys. J.
  C} {\bf 82} (2022) 597}, [\href{http://arxiv.org/abs/2112.05163}{{\tt
  2112.05163}}].

\bibitem{CMS:2023mny}
{\scshape CMS} Collaboration, A.~Hayrapetyan et~al., \emph{{Search for
  supersymmetry in final states with disappearing tracks in proton-proton
  collisions at $\sqrt{s}$ = 13 TeV}},
  \href{http://arxiv.org/abs/2309.16823}{{\tt 2309.16823}}.

\bibitem{Giudice:1998xp}
G.~F. Giudice, M.~A. Luty, H.~Murayama and R.~Rattazzi, \emph{{Gaugino mass
  without singlets}},
  \href{http://dx.doi.org/10.1088/1126-6708/1998/12/027}{\emph{J. High Energy
  Phys.} {\bf 12} (1998) 027}, [\href{http://arxiv.org/abs/hep-ph/9810442}{{\tt
  hep-ph/9810442}}].

\bibitem{Randall:1998uk}
L.~Randall and R.~Sundrum, \emph{{Out of this world supersymmetry breaking}},
  \href{http://dx.doi.org/10.1016/S0550-3213(99)00359-4}{\emph{Nucl. Phys. B}
  {\bf 557} (1999) 79--118}, [\href{http://arxiv.org/abs/hep-th/9810155}{{\tt
  hep-th/9810155}}].

\bibitem{Baer:2011ec}
H.~Baer, V.~Barger and P.~Huang, \emph{{Hidden SUSY at the LHC: the light
  higgsino-world scenario and the role of a lepton collider}},
  \href{http://dx.doi.org/10.1007/JHEP11(2011)031}{\emph{J. High Energy Phys.}
  {\bf 11} (2011) 031}, [\href{http://arxiv.org/abs/1107.5581}{{\tt
  1107.5581}}].

\bibitem{Nagata:2014wma}
N.~Nagata and S.~Shirai, \emph{{Higgsino Dark Matter in High-Scale
  Supersymmetry}}, \href{http://dx.doi.org/10.1007/JHEP01(2015)029}{\emph{J.
  High Energy Phys.} {\bf 01} (2015) 029},
  [\href{http://arxiv.org/abs/1410.4549}{{\tt 1410.4549}}].

\bibitem{Baer:2015tva}
H.~Baer, V.~Barger, P.~Huang, D.~Mickelson, M.~Padeffke-Kirkland and X.~Tata,
  \emph{{Natural SUSY with a bino- or wino-like LSP}},
  \href{http://dx.doi.org/10.1103/PhysRevD.91.075005}{\emph{Phys. Rev. D} {\bf
  91} (2015) 075005}, [\href{http://arxiv.org/abs/1501.06357}{{\tt
  1501.06357}}].

\bibitem{ATLAS:2021moa}
{\scshape ATLAS} Collaboration, G.~Aad et~al., \emph{{Search for
  chargino\textendash{}neutralino pair production in final states with three
  leptons and missing transverse momentum in $\sqrt{s} = 13$~TeV pp collisions
  with the ATLAS detector}},
  \href{http://dx.doi.org/10.1140/epjc/s10052-021-09749-7}{\emph{Eur. Phys. J.
  C} {\bf 81} (2021) 1118}, [\href{http://arxiv.org/abs/2106.01676}{{\tt
  2106.01676}}].

\bibitem{ATLAS:2019lng}
{\scshape ATLAS} Collaboration, G.~Aad et~al., \emph{{Searches for electroweak
  production of supersymmetric particles with compressed mass spectra in
  $\sqrt{s}=$ 13 TeV $pp$ collisions with the ATLAS detector}},
  \href{http://dx.doi.org/10.1103/PhysRevD.101.052005}{\emph{Phys. Rev. D} {\bf
  101} (2020) 052005}, [\href{http://arxiv.org/abs/1911.12606}{{\tt
  1911.12606}}].

\bibitem{CMS:2023qhl}
{\scshape CMS} Collaboration, A.~Tumasyan et~al., \emph{{Combined search for
  electroweak production of winos, binos, higgsinos, and sleptons in
  proton-proton collisions at $sqrt{s}=$ 13 TeV}}, {\emph{CMS-PAS-SUS-21-008}
  (2023) }.

\bibitem{CMS:2021edw}
{\scshape CMS} Collaboration, A.~Tumasyan et~al., \emph{{Search for
  supersymmetry in final states with two or three soft leptons and missing
  transverse momentum in proton-proton collisions at $ \sqrt{s} $ = 13 TeV}},
  \href{http://dx.doi.org/10.1007/JHEP04(2022)091}{\emph{J. High Energy Phys.}
  {\bf 04} (2022) 091}, [\href{http://arxiv.org/abs/2111.06296}{{\tt
  2111.06296}}].

\bibitem{CMS:2021cox}
{\scshape CMS} Collaboration, A.~Tumasyan et~al., \emph{{Search for electroweak
  production of charginos and neutralinos in proton-proton collisions at $
  \sqrt{s} $ = 13 TeV}},
  \href{http://dx.doi.org/10.1007/JHEP04(2022)147}{\emph{J. High Energy Phys.}
  {\bf 04} (2022) 147}, [\href{http://arxiv.org/abs/2106.14246}{{\tt
  2106.14246}}].

\bibitem{LEPlimits}
{ALEPH, DELPHI, L3, OPAL Experiments}, \emph{{Combined LEP Chargino Results, up
  to 208 GeV for low DM}}, {\emph{{LEPSUSYWG/02-04.1}} (2002) }.

\bibitem{ATLAS:2021kxv}
{\scshape ATLAS} Collaboration, G.~Aad et~al., \emph{{Search for new phenomena
  in events with an energetic jet and missing transverse momentum in $pp$
  collisions at $\sqrt {s}$ =13 TeV with the ATLAS detector}},
  \href{http://dx.doi.org/10.1103/PhysRevD.103.112006}{\emph{Phys. Rev. D} {\bf
  103} (2021) 112006}, [\href{http://arxiv.org/abs/2102.10874}{{\tt
  2102.10874}}].

\bibitem{CMS:2021far}
{\scshape CMS} Collaboration, A.~Tumasyan et~al., \emph{{Search for new
  particles in events with energetic jets and large missing transverse momentum
  in proton-proton collisions at $ \sqrt{s} $ = 13 TeV}},
  \href{http://dx.doi.org/10.1007/JHEP11(2021)153}{\emph{J. High Energy Phys.}
  {\bf 11} (2021) 153}, [\href{http://arxiv.org/abs/2107.13021}{{\tt
  2107.13021}}].

\bibitem{PhysRevD.37.2515}
J.~F. Gunion and H.~E. Haber, \emph{Two-body decays of neutralinos and
  charginos}, \href{http://dx.doi.org/10.1103/PhysRevD.37.2515}{\emph{Phys.
  Rev. D} {\bf 37} (May, 1988) 2515--2532}.

\bibitem{Dumont:2014tja}
B.~Dumont, B.~Fuks, S.~Kraml, S.~Bein, G.~Chalons, E.~Conte et~al.,
  \emph{{Toward a public analysis database for LHC new physics searches using
  MADANALYSIS 5}},
  \href{http://dx.doi.org/10.1140/epjc/s10052-014-3242-3}{\emph{Eur. Phys. J.}
  {\bf C75} (2015) 56}, [\href{http://arxiv.org/abs/1407.3278}{{\tt
  1407.3278}}].

\bibitem{Conte:2018vmg}
E.~Conte and B.~Fuks, \emph{{Confronting new physics theories to LHC data with
  MADANALYSIS 5}},
  \href{http://dx.doi.org/10.1142/S0217751X18300272}{\emph{Int. J. Mod. Phys.}
  {\bf A33} (2018) 1830027}, [\href{http://arxiv.org/abs/1808.00480}{{\tt
  1808.00480}}].

\bibitem{DVN/IRF7ZL_2021}
A.~Albert, \emph{{Implementation of a search for new phenomena in events
  featuring energetic jets and missing transverse energy (137 fb-1; 13 TeV;
  CMS-EXO-20-004)}},
  \href{http://dx.doi.org/10.14428/DVN/IRF7ZL}{\emph{10.14428/DVN/IRF7ZL}
  (2021) }.

\bibitem{DVN/REPAMM_2023}
D.~Agin, \emph{{Implementation of a search for new physics with jets and
  missing transverse energy (139/fb; 13 TeV; ATLAS-EXOT-2018-06)}},
  \href{http://dx.doi.org/10.14428/DVN/REPAMM}{\emph{10.14428/DVN/REPAMM}
  (2023) }.

\bibitem{Alwall:2014hca}
J.~Alwall, R.~Frederix, S.~Frixione, V.~Hirschi, F.~Maltoni, O.~Mattelaer
  et~al., \emph{{The automated computation of tree-level and next-to-leading
  order differential cross sections, and their matching to parton shower
  simulations}}, \href{http://dx.doi.org/10.1007/JHEP07(2014)079}{\emph{J. High
  Energy Phys.} {\bf 07} (2014) 079},
  [\href{http://arxiv.org/abs/1405.0301}{{\tt 1405.0301}}].

\bibitem{Darme:2023jdn}
L.~Darm\'e et~al., \emph{{UFO 2.0: the \textquoteleft{}Universal Feynman
  Output\textquoteright{} format}},
  \href{http://dx.doi.org/10.1140/epjc/s10052-023-11780-9}{\emph{Eur. Phys. J.
  C} {\bf 83} (2023) 631}, [\href{http://arxiv.org/abs/2304.09883}{{\tt
  2304.09883}}].

\bibitem{Duhr:2011se}
C.~Duhr and B.~Fuks, \emph{{A superspace module for the FeynRules package}},
  \href{http://dx.doi.org/10.1016/j.cpc.2011.06.009}{\emph{Comput. Phys.
  Commun.} {\bf 182} (2011) 2404--2426},
  [\href{http://arxiv.org/abs/1102.4191}{{\tt 1102.4191}}].

\bibitem{NNPDF:2021njg}
{\scshape NNPDF} Collaboration, R.~D. Ball et~al., \emph{{The path to proton
  structure at 1\% accuracy}},
  \href{http://dx.doi.org/10.1140/epjc/s10052-022-10328-7}{\emph{Eur. Phys. J.
  C} {\bf 82} (2022) 428}, [\href{http://arxiv.org/abs/2109.02653}{{\tt
  2109.02653}}].

\bibitem{Buckley:2014ana}
A.~Buckley, J.~Ferrando, S.~Lloyd, K.~Nordstr\"om, B.~Page, M.~R\"ufenacht
  et~al., \emph{{LHAPDF6: parton density access in the LHC precision era}},
  \href{http://dx.doi.org/10.1140/epjc/s10052-015-3318-8}{\emph{Eur. Phys. J.
  C} {\bf 75} (2015) 132}, [\href{http://arxiv.org/abs/1412.7420}{{\tt
  1412.7420}}].

\bibitem{Sjostrand:2014zea}
T.~Sjöstrand, S.~Ask, J.~R. Christiansen, R.~Corke, N.~Desai, P.~Ilten et~al.,
  \emph{{An Introduction to PYTHIA 8.2}},
  \href{http://dx.doi.org/10.1016/j.cpc.2015.01.024}{\emph{Comput. Phys.
  Commun.} {\bf 191} (2015) 159--177},
  [\href{http://arxiv.org/abs/1410.3012}{{\tt 1410.3012}}].

\bibitem{Mangano:2006rw}
M.~L. Mangano, M.~Moretti, F.~Piccinini and M.~Treccani, \emph{{Matching matrix
  elements and shower evolution for top-quark production in hadronic
  collisions}}, \href{http://dx.doi.org/10.1088/1126-6708/2007/01/013}{\emph{J.
  High Energy Phys.} {\bf 01} (2007) 013},
  [\href{http://arxiv.org/abs/hep-ph/0611129}{{\tt hep-ph/0611129}}].

\bibitem{Alwall:2008qv}
J.~Alwall, S.~de~Visscher and F.~Maltoni, \emph{{QCD radiation in the
  production of heavy colored particles at the LHC}},
  \href{http://dx.doi.org/10.1088/1126-6708/2009/02/017}{\emph{J. High Energy
  Phys.} {\bf 02} (2009) 017}, [\href{http://arxiv.org/abs/0810.5350}{{\tt
  0810.5350}}].

\bibitem{Araz:2020lnp}
J.~Y. Araz, B.~Fuks and G.~Polykratis, \emph{{Simplified fast detector
  simulation in MADANALYSIS 5}},
  \href{http://dx.doi.org/10.1140/epjc/s10052-021-09052-5}{\emph{Eur. Phys. J.
  C} {\bf 81} (2021) 329}, [\href{http://arxiv.org/abs/2006.09387}{{\tt
  2006.09387}}].

\bibitem{deFavereau:2013fsa}
{\scshape DELPHES 3} Collaboration, J.~de~Favereau, C.~Delaere, P.~Demin,
  A.~Giammanco, V.~Lemaître, A.~Mertens et~al., \emph{{DELPHES 3, A modular
  framework for fast simulation of a generic collider experiment}},
  \href{http://dx.doi.org/10.1007/JHEP02(2014)057}{\emph{J. High Energy Phys.}
  {\bf 02} (2014) 057}, [\href{http://arxiv.org/abs/1307.6346}{{\tt
  1307.6346}}].

\bibitem{Fuks:2013vua}
B.~Fuks, M.~Klasen, D.~R. Lamprea and M.~Rothering, \emph{{Precision
  predictions for electroweak superpartner production at hadron colliders with
  {\sc Resummino}}},
  \href{http://dx.doi.org/10.1140/epjc/s10052-013-2480-0}{\emph{Eur. Phys. J.
  C} {\bf 73} (2013) 2480}, [\href{http://arxiv.org/abs/1304.0790}{{\tt
  1304.0790}}].

\bibitem{Fiaschi:2023tkq}
J.~Fiaschi, B.~Fuks, M.~Klasen and A.~Neuwirth, \emph{{Electroweak superpartner
  production at 13.6 Tev with Resummino}},
  \href{http://dx.doi.org/10.1140/epjc/s10052-023-11888-y}{\emph{Eur. Phys. J.
  C} {\bf 83} (2023) 707}, [\href{http://arxiv.org/abs/2304.11915}{{\tt
  2304.11915}}].

\bibitem{Read:2002hq}
A.~L. Read, \emph{{Presentation of search results: The $CL_s$ technique}},
  \href{http://dx.doi.org/10.1088/0954-3899/28/10/313}{\emph{J. Phys. G} {\bf
  28} (2002) 2693--2704}.

\bibitem{Araz:2023bwx}
J.~Y. Araz, \emph{{Spey: smooth inference for reinterpretation studies}},
  \href{http://arxiv.org/abs/2307.06996}{{\tt 2307.06996}}.

\bibitem{Cowan:2010js}
G.~Cowan, K.~Cranmer, E.~Gross and O.~Vitells, \emph{{Asymptotic formulae for
  likelihood-based tests of new physics}},
  \href{http://dx.doi.org/10.1140/epjc/s10052-011-1554-0}{\emph{Eur. Phys. J.
  C} {\bf 71} (2011) 1554}, [\href{http://arxiv.org/abs/1007.1727}{{\tt
  1007.1727}}].

\bibitem{Heinrich:2021gyp}
L.~Heinrich, M.~Feickert, G.~Stark and K.~Cranmer, \emph{{pyhf: pure-Python
  implementation of HistFactory statistical models}},
  \href{http://dx.doi.org/10.21105/joss.02823}{\emph{J. Open Source Softw.}
  {\bf 6} (2021) 2823}.

\bibitem{Thomas:1998wy}
S.~D. Thomas and J.~D. Wells, \emph{{Phenomenology of Massive Vectorlike
  Doublet Leptons}},
  \href{http://dx.doi.org/10.1103/PhysRevLett.81.34}{\emph{Phys. Rev. Lett.}
  {\bf 81} (1998) 34--37}, [\href{http://arxiv.org/abs/hep-ph/9804359}{{\tt
  hep-ph/9804359}}].

\bibitem{Porod:2003um}
W.~Porod, \emph{{SPheno, a program for calculating supersymmetric spectra, SUSY
  particle decays and SUSY particle production at e+ e- colliders}},
  \href{http://dx.doi.org/10.1016/S0010-4655(03)00222-4}{\emph{Comput. Phys.
  Commun.} {\bf 153} (2003) 275--315},
  [\href{http://arxiv.org/abs/hep-ph/0301101}{{\tt hep-ph/0301101}}].

\bibitem{Porod:2011nf}
W.~Porod and F.~Staub, \emph{{SPheno 3.1: Extensions including flavour,
  CP-phases and models beyond the MSSM}},
  \href{http://dx.doi.org/10.1016/j.cpc.2012.05.021}{\emph{Comput. Phys.
  Commun.} {\bf 183} (2012) 2458--2469},
  [\href{http://arxiv.org/abs/1104.1573}{{\tt 1104.1573}}].

\bibitem{Staub:2017jnp}
F.~Staub and W.~Porod, \emph{{Improved predictions for intermediate and heavy
  Supersymmetry in the MSSM and beyond}},
  \href{http://dx.doi.org/10.1140/epjc/s10052-017-4893-7}{\emph{Eur. Phys. J.
  C} {\bf 77} (2017) 338}, [\href{http://arxiv.org/abs/1703.03267}{{\tt
  1703.03267}}].

\bibitem{ATLAS:2018nda}
{\scshape ATLAS} Collaboration, M.~Aaboud et~al., \emph{{Search for dark matter
  in events with a hadronically decaying vector boson and missing transverse
  momentum in $pp$ collisions at $\sqrt{s} = 13$ TeV with the ATLAS detector}},
  \href{http://dx.doi.org/10.1007/JHEP10(2018)180}{\emph{J. High Energy Phys.}
  {\bf 10} (2018) 180}, [\href{http://arxiv.org/abs/1807.11471}{{\tt
  1807.11471}}].

\bibitem{Carpenter:2023agq}
L.~M. Carpenter, H.~Gilmer, J.~Kawamura and T.~Murphy, \emph{{Taking aim at the
  wino-Higgsino plane with the LHC}},
  \href{http://dx.doi.org/10.1103/PhysRevD.109.015012}{\emph{Phys. Rev. D} {\bf
  109} (2024) 015012}, [\href{http://arxiv.org/abs/2309.07213}{{\tt
  2309.07213}}].

\bibitem{Fuks:2017rio}
B.~Fuks, M.~Klasen, S.~Schmiemann and M.~Sunder, \emph{{Realistic simplified
  gaugino-higgsino models in the MSSM}},
  \href{http://dx.doi.org/10.1140/epjc/s10052-018-5695-2}{\emph{Eur. Phys. J.
  C} {\bf 78} (2018) 209}, [\href{http://arxiv.org/abs/1710.09941}{{\tt
  1710.09941}}].

\bibitem{Waltenberger:2016vxp}
{\scshape SModelS} Collaboration, W.~Waltenberger, \emph{{SModelS: A Tool for
  Making Systematic Use of Simplified Models Results}},
  \href{http://dx.doi.org/10.1088/1742-6596/762/1/012076}{\emph{J. Phys. Conf.
  Ser.} {\bf 762} (2016) 012076}.

\bibitem{Alguero:2021dig}
G.~Alguero, J.~Heisig, C.~K. Khosa, S.~Kraml, S.~Kulkarni, A.~Lessa et~al.,
  \emph{{Constraining new physics with SModelS version 2}},
  \href{http://dx.doi.org/10.1007/JHEP08(2022)068}{\emph{J. High Energy Phys.}
  {\bf 08} (2022) 068}, [\href{http://arxiv.org/abs/2112.00769}{{\tt
  2112.00769}}].

\bibitem{MahdiAltakach:2023bdn}
M.~Mahdi~Altakach, S.~Kraml, A.~Lessa, S.~Narasimha, T.~Pascal and
  W.~Waltenberger, \emph{{SModelS v2.3: enabling global likelihood analyses}},
  \href{http://dx.doi.org/10.21468/SciPostPhys.15.5.185}{\emph{SciPost Phys.}
  {\bf 15} (2023) 185}, [\href{http://arxiv.org/abs/2306.17676}{{\tt
  2306.17676}}].

\end{thebibliography}\endgroup

\end{document}